\documentstyle[12pt]{article}
\begin{document}
\title{PECULIARITIES OF RAMAN SPECTRA SHAPE IN THE DISORDERED
FERROELECTRICS}

\author{M.D.~Glinchuk, I.V.~Kondakova}
\maketitle

\begin{center}
\it{Institute for Material Sciences, National Academy of Sciences of Ukraine,
\\
Krzhizhanovskogo str. 3, 252180 Kiev, UKRAINE}
\end{center}
\bigskip

\begin{abstract}
Theory of the first order Raman scattering (FOR) line shape allowing for
nonlinear and correlation effects contribution to inhomogeneous broadening
as well as dynamic mechanisms of homogeneous broadening is developed. It is
shown that in general case FOR line contains two maxima, shape of low
frequency one being defined by homogeneous broadening. Our theory explains the
peculiarity of observed FOR scattering of $TO_{2}$ hard phonon in $KTL$
and $KTN$ at different temperatures and concentrations of $Li$ and  $Nb $ ions.

\end{abstract}


\section{Introduction}

Investigation of Raman spectra of the first order (FOR) in disordered
ferroelectrics was shown to be sensitive method for studying of the critical
slowing down of optic phonons nearby ferroelectric transition and dynamics
of local fluctuation (see e.g. \cite{1,2} and ref. therein). Several
attempts were made to explain FOR spectral anomalies such as strong increase
of peak intensity in the vicinity of $T_{c}$ and peculiar shape of the line
\cite{3} . However, up to now there is no general approach to the description 
of Raman spectra in the disordered ferroelectrics nearby $T_{c}$. These results 
are in the discrepancy in the values of the materials physical characteristics
 obtained from experimental data (see e.g. \cite{1} -\cite{4}). 
The most promising
way for development of general approach seems to be the model allowing for
inhomogeneous broadening of FOR lines induced by static disorder and
homogeneous broadening due to dynamic effects. The existence of both these 
contributions to Raman spectra
was shown early in $KTa_{1-x}Nb_{x}O_{3}$ ($ KTN$) and $K_{1-x}Li_{x}TaO_{3}$
($KLT$) \cite{1,2,3}.

Theory of inhomogeneously broadened lines in radio-, optic- and other
spectroscopy methods was developed in many details early for conventional
dielectric and magnetic systems \cite{5,6}. In these materials inhomogeneous
broadening used to be temperature independent, so that only homogeneous
broadening may be the source of temperature dependence of the spectra line shape 
and width. In the disordered ferroelectrics due to nonlinear and spatial 
correlation effects which is known to be especially strong in the vicinity of 
$T_{c}$ inhomogeneous broadening was recently shown to be temperature dependent 
\cite{glinchuk98} .

In the present work a theory of FOR allowing for both linear and nonlinear
contributions to inhomogeneous broadening as well as to homogeneous mechanisms
of line broadening is developed. The theory  explains the
main peculiarities of observed FOR spectra in $KTL$ with 1\% and 4\% of
$Li $ \cite{3} and in $KTN$ with 15.7\% of $Nb $ ions \cite{2} when the systems 
approach the phase transition from above. 
Since in vicinity of the phase transition
the nonlinear effects are known to be large enough our theory is more general
than that in \cite{2,3} , where nonlinear and spatial correlation effects were
not taken into account. The proposed theoretical approach allows to 
describe of the Raman spectra (in distingvish \cite{2,3} ) without 
calculation of the correlation function of the quasi-static polarization 
fluctuations but only suggesting about its spectral density form. Because of 
generality of proposed theory it can be applied to many ferroelectrics in 
vicinity of transition temperature.


\section{Theory of first order Raman spectra shape}

2.1 Incipient ferroelectric $KTaO_{3}$ doped by $Li$ or $Nb$ ions can have
ferroelectric phase transition, mixed ferroglass phase and dipole glass state in
dependence on temperature and the impurities concentrations \cite{Gl,St}. These
materials can be regarded as the model systems  of the disordered ferroelectrics. 
Their
peculiar properties are connected with off-center positions of $Li$ and $Nb$
ions substituted for $K^{+}$ and $Ta^{5+}$ ions respectively. Some anomalies of
Raman spectra can be explained also by these ions off-centrality. In particular
quasistatic polarization fluctuations induced by off-center ions can lead to the
reduction of $KTaO_{3}$ cubic symmetry and to the appearance of first order
Raman scattering (FOR) even above the transition temperature $T_{c}$ \cite{1,3}.
The most detail description of the appeareance of the single phonon Raman lines 
above $T_{c}$ can be found in \cite{bruce} where it was shown how the slow 
finite-range precursor order, dominant near the phase transition, is responsible 
for the persistence of quasi-first-order hard-mode features above $T_{c}$.

As known, Raman scattering in perovskite ferroelectrics is caused by the changes
in the oxygen electrtonic polarizability $\delta \alpha (r, t)$ induced by 
optic vibration
modes of the lattice. This changes can be written in the form
\begin{equation}
  \delta \alpha (r, t)= \mathbf{P} (\mathbf{r} , t) \cdot \hat{\Lambda } \cdot 
  \mathbf{P} (\mathbf{r} , t)
  \label{1}
\end{equation}
where $P(r, t)$ represents the space- and time-dependent polarization
fluctuation and $ \hat{\Lambda } $ is a fourth-rank tensor. Equation (1) 
usually describes second-order Raman scattering. 
However, if we write the fluctuation polarization as a sum
over the components due to the polar hard modes $P^{h}$  and slow relaxing 
component
$P^{\mu }$ we can explain the appearance of single-phonon Raman scattering above
$T_{c}$ considering the cross terms $P^{\mu } P^{h}$ in the Eq.(1)
The scattering intensity  is given by the spatial and temperature 
Fourier components of the polarizability correlation function
\begin{eqnarray}    
I(\omega )& \sim  &<\delta \alpha (r,t) \delta \alpha (0,0)>_{q=0,\omega }
\nonumber  \\ 
& \sim  & \sum_{q'} \int d\omega '<P^{\mu} (r,t) P^{\mu } (0,0)>_{q',\omega '}\, 
<P^h (r,t) P^h (0,0)>_{-q',\omega -\omega '} \, , 
\label{2} 
\end{eqnarray}
Because the first correlation function has a sharp maximum near $\omega '= 0$ 
and second one near $\Omega _{q}$, $I(\omega )$ is sharply peaked at 
$\omega  = \Omega _{q}$, i.e., near the hard single-phonon frequency.

It is seen that (\ref{2}) can be represented as the convolution of
the functions
represented inhomogeneous and homogeneous mechanisms of broadening connected
respectively with the first and the second correlators and so it can be
written in the form
\begin{equation}
I(\omega )=I_{0} \int_{-\infty }^{\infty }
 J(\omega ,\omega ')\,
f(\omega ^{\prime}) d\omega ^{\prime}
\ .
\label{3}
\end{equation}
The detailed form of
homogeneuos $J(\omega ,\omega ^{\prime} )$ and inhomogeneous contribution
$f(\omega ^{\prime })$ can be obtained by the following way. Thermal lattice
vibrations usually lead to homogeneous broadening with simple Lorentzian shape,
i.e.
\begin{equation}
<P^h (r,t) P^h (0,0)>_{-q',\omega -\omega '} \sim 
- \frac{1}{\Omega _{q'}} \,
\frac{\Gamma }
   {\Gamma ^2+\left(\omega -\omega ^{\prime}-\Omega _{q^{\prime}}\right)^2}
\, ,
\label{4}
\end{equation}
which should be valid for the difference of frequencies $(\omega -\omega ')$ close
to hard phonon frequency $\Omega _{q'}$. 
On the other hand the correlation function of quasistatic polarization reads
\begin{equation}
<P^\mu (r,t) P^\mu (0,0)>_{q',\omega '}=
<P^{\mu} (r) P^{\mu } (0)>_{q^{\prime}} \, \pi  \delta (\omega ')
\, .
\label{5}
\end{equation}
Substituting (long-wave approximation) 
$\sum_{q^{\prime}} \rightarrow \frac{V}{(2\pi )^3} \int d^3q'$ 
and performing some transformation one obtaine that
\begin{equation}
f(\omega ') =
\int dq' \, q'^2<P^\mu (r) P^\mu (0)>_{q'}  \delta (\omega '-\Omega _{q'})
\frac{1}{\Omega_{q'}}
\, ,
\label{6}
\end{equation}
\begin{equation}
 J(\omega ,\omega ^{\prime})=
 \frac{\Gamma }{\Gamma ^2 + (\omega -\omega '^2)},
 \ \ \ \
  I_0=\frac{V}{2\pi }
\, .
\label{7}
\end{equation}
Equations (\ref{6}) and (\ref{7}) determine respectively the contribution of
inhomogeneous and homogeneous broadening into Raman line shape
(see Eq.~(\ref{3})). However it is seen from Eq.~(\ref{6}),
that for $f(\omega )$ calculation
one has to calculate firstly the correlation function of polarization. Its
calculation utilized some special models and supposition \cite{1,3} . To our
mind more general approach for $f(\omega )$ calculation can be statistical
theory of specrtal line shape \cite{5}. This theory was sucsessfully applied for
description of optic, radiospectrroscopy, $\gamma $-resonance etc. line
shapes \cite{6,glinchuk97}.

2.2 The shape of $f(\omega )$ was calculated in the
statistical theory framework for the cases when the frequency shift of spectral
line is linear \cite{5,6} and nonlinear \cite{glinchuk98,glinchuk97} function of 
random fields.
The latter case seems to be especially important for disordered ferroelectrics
at $T=T_{c}\pm \Delta T$, $\Delta T\simeq (10\div 20) K$, where nonlinear and
correlation effects is known to be large enough. As a matter of fact the 
distribution function 
allowing for these effects can be expressed via linear one in the framework of 
the general 
theory of probability which makes it possible to write down the distribution of 
one random 
quantity via that of another if the
relation between these quantities is known \cite{hudson}. For example in the
simplest case when the random quantity $x$ is a single-valued monotonous
function of another random quantity $h$ then the relation between their
distribution functions $g(h)$ and $f(x)$ can be written as \cite{hudson} 
\begin{equation}
g(h)=f(x(h))\left| \frac{dx(h)}{dh} \right|
\, .
\label{8}
\end{equation}
In general case when several different $x$ values correspond to the same
$h(x)$ value the space of $x$ should be devided into the regions where the
function $h(x)$ is monotonous. For the entire $x$-domain distribution
function $g(h)$ can be represented as a sum of terms like Eq.(8)
\cite{glinchuk98,glinchuk97,hudson}. Since the distribution function in the
case of linear contribution of the random field $f_{1}(\omega )$ can be
analytically calculated in statistical theory approach \cite{5,6} we have to
express via it the distribution function allowing for nonlinear random field
contribution. Let us suppose that the shift
$\Delta \omega =\omega-\omega _{0} \equiv \omega $
of the spectral line maximum position 
$\omega {0}$ 
due to random field contribution can be written as a power law (up to some
$m^{th}$ power) of the random field $\omega '$
\begin{equation}
\omega =\omega
'-\alpha _{2}\omega '^2- \dots - \alpha _m\omega '^m 
\, .
\label{9} 
\end{equation} 
Equation (9) makes it possible to express the distribution function of 
$\omega $ via
that of $\omega '$ in the form \cite{glinchuk98,glinchuk97,hudson}
\begin{equation}
f_m(\omega ) = 
\sum_{k=1}^m f_1\left( \omega '={\omega }_k\right)  \left| 
\frac{d\phi (\omega ,\omega ^{\prime})}{d\omega ^{\prime }}
\right| _{\omega ^{\prime }=\omega _k} \, ,
\label{10}
\end{equation}
\begin{equation} 
\label{11}
\phi (\omega ,\omega^\prime ) =
\omega -\omega ^{\prime}-\alpha _{2}{\omega ^{\prime}}^{2}- \dots
-\alpha _{m}{\omega ^\prime }^{m} \, .  
\end{equation}
where $\omega _{k}$ are the real roots of the algebraic equation
\begin{equation} 
\label{12} 
\phi (\omega ,\omega _{k}) = 0 \, .
\end{equation} 
It is seen, that $m^{th}$ order
distribution function $f_{m}(\omega )$  is expressed via that calculated in
linear approximation when all nonlinear coefficients equal zero, i.e. 
$\alpha _2=\alpha _3=\dots =\alpha _m=0$.  
Shape of $f_{1}(\omega )$ calculated in the
statistical theory approach was shown to be Gaussian, Lorentzian or Holtzmarkian
in dependence on types of random field sources with parameters determined by the
sources concentrations and characteristics \cite{St,6}.

In the simplest case, when the main contribution is connected with the first
nonlinear term in Eqs.~(\ref{9}), (\ref{11})
($\alpha _2\neq 0, \alpha _{3}=\dots =\alpha_{m}=0$)
Eqs.(10), (11) and (12) lead to the following form of the
normalized second order distribution function
\begin{equation}
\label{13}
f_2(\omega )=\frac{\Theta \left( \omega +\frac{1}{4\alpha _2}
\right)} {\sqrt{1+4\alpha _2\omega } } \left[ f_1\left( \frac{\sqrt{
1+4\alpha _2\omega } -1}{2\alpha _2} \right) +f_1\left( -\frac{\sqrt{
1+4\alpha _2\omega } +1}{2\alpha _2} \right) \right]
\, ,
\end{equation}
\noindent
where $\Theta $ is the teta-function, so that $f_{2}\neq 0$ only in region 
$\omega _{c}\leq \omega \leq \infty $ ($\alpha _{2}>0$) or 
$-\infty \leq \omega \leq \omega _{c}$ ($\alpha _{2}<0$),
\begin{equation}
\label{14}\omega _c=-(1/4\alpha _2)
\end{equation}
\noindent
is a critical frequency at which a divergency of $f_{2}(\omega )$ appears.
Therefore $f_{2}(\omega )$ is strongly asymmetrical, and its form at 
$\omega \gg  \omega _{c}$ (the wing of $f_{2}(\omega )$) 
is the following
\begin{equation}
\label{15} 
f_{2}(\omega )\rightarrow 1/\sqrt{1+4\alpha _{2}\omega } 
\, .
\end{equation}
Let us consider the homogeneous contribution in the Lorentzian
form (see Eq.(7)). In this case the integration in Eq.(\ref{3}) is equivalent
to substitution $\omega \pm \frac{i}{\tau }$  for $\omega $ in the
Eq.(\ref{13})
($1/\tau \equiv \Gamma $ is half width on the half height). 
If there are several mechanisms of homogeneous broadening with Lorentzian forms
$1/\tau =\sum _{i}1/\tau _{i}$, where $i$ numerates the mechanisms.

The aforementioned procedure (or integration in Eq.(\ref{3} )) with respect to
Eq.(\ref{13}) in supposition that $f_{1}(\omega )$ has Gaussian form, i.e.
\begin{equation}
\label{16}
f_{1}(\omega ) = 
\frac{1}{\sqrt{2\pi } } e^{\textstyle -\frac{\omega ^{2}}{2\Delta ^{2}} }
\, ,
\end{equation}
\noindent
leads to the following shape of spectral line in the considered case
\begin{eqnarray}
I_2(\omega ) =
\frac{ \frac{1}{2}+\frac{1}{\pi } \arctan {\tau (\omega
-\omega _{c})} }{\Delta \sqrt{2\pi \varphi (\omega ) }} 
\left\{ \exp \left[
\frac{S_{1}(\omega )-2(1+2\alpha _{2}\omega )}{8\alpha _{2}^{2}\Delta ^{2}}
 \right] \cos{\frac{4\alpha _{2}/\tau -S_{2}(\omega )}{8\alpha _{2}^{2}\Delta
^{2}} } \right.
\nonumber  \\
+ \left. \exp \left[\frac{-S_{1}(\omega )-2(1+2\alpha
 _{2}\omega )}{8\alpha _{2}^{2}\Delta ^{2}} \right] \cos{\frac{4\alpha _{2}/\tau
 +S_{2}(\omega )}{8\alpha _{2}^{2}\Delta ^{2}} } \right\} \, ,
 \label{17}
 \end{eqnarray}
where
\begin{eqnarray*}
\varphi (\omega
)=\sqrt{(1+4\alpha _{2}\omega )^{2}+(4\alpha _{2}/\tau )^{2}} \, ,
\nonumber  \\
S_{1,2}(\omega
 )=\sqrt{2} \sqrt{\varphi (\omega )\pm (1+4\alpha _{2}\omega )} \, .
 \nonumber
\end{eqnarray*}
The results of numerical calculations for several values of dimensionless
parameters $\alpha _{2}\Delta $ and $1/(\tau \Delta )$ are depicted in 
figs.~1,2. It is seen that homogeneous contribution transforms $f_{2}(\omega )$
divergence at $\omega =\omega _{c}$ into sharp maximum, so that the spectral
line has two maxima (see fig.~1) instead of one in the linear case origin of
high frequency one being connected with Gaussian form. The
distance  between two maxima approximately equals $\omega _{c}$ (see fig.~1).
At large  enough nonlinear contribution only sharp maximum at $\omega
=\omega _{c}$  conserves, its width increases with $1/\tau $ increasing (see
fig.~2). The left  hand side half-width of low-frequency peak completely
defined by homogeneous  broadening mechanisms, meanwhile right hand side
part of the line at $\omega  >\omega _{c}$ defines mainly by inhomogeneous
mechanism contribution.

\section{Raman spectra in KTL and KTN}

3.1 The developed theory was applied to recently observed $TO_2$ FOR in $KTL$
with 1\% and 4\% of $Li$ and in $KTN$ with  15.7\% of $Nb$ \cite{3,2}.

Let us begin with the consideration of Raman spectra in $KTL$. Measurements were
 carried out at $T=10$~K ($ x_{Li} =0.01$) and $T=55$~K ($x_{Li} = 0.04$) \cite{3}. 
In both samples the observed line was strongly asymmetric with maximum at
$\omega \simeq 198~{\rm cm}^{-1}$. 
These lines were fitted good enough by Eqs. (12) - (14) with the following 
dimensionless parameters
 $\alpha _{2}\Delta =0.38$, $(\tau \Delta )^{-1} =0.06$ (fig.~3) and
 $\alpha _{2}\Delta =0.55$, $(\tau \Delta )^{-1} =0.15$ (fig.~4)
for the considered $Li$ concentrations respectively.

 The position of the lines maxima  $\omega _{m}-\omega _{0}=\omega _{c}$ is
defined by the parameter of nonlinearity in accordance with Eq.~(\ref{14}). 
We obtained
$\alpha _{2}\simeq 0.1 $~cm 
by the fitting of the high frequency line tails with the Eq.(15). 
This value gives 
$\omega _{m}=\omega _{c}+\omega _{0} \simeq 197.6~{\rm cm}^{-1}$ 
with the reference point $\omega _{0} = 199~{\rm cm}^{-1}$. Note that Eq.(15)
describes also observed frequency dependence of the lines tails for the same
$\alpha _{2}$ value. This made it possible to obtaine yhe magnitude of the
Gaussian width $\Delta $ with the help of aforemationed $\alpha _{2}\Delta $
values: $ \Delta =3.8~{\rm cm}^{-1}$ ($x_{Li}=0.01$) and 
$\Delta = 5.5~{\rm cm}^{-1}$
($x_{Li}=0.04)$.

In accordance with our theory the line form at $\omega < \omega _{c}$ defines
 completely by homogeneous mechanism contribution represented by Eq.(4). It is
seen that low-frequency half-width $1/\tau $ is connected with hard phonon life
time $\Gamma ^{-1}$. Keeping in mind the obtained values of $(\tau \Delta
)^{-1} \equiv \Gamma /\Delta $ and $\Delta $ one find $\Gamma $ ($T=10$~K) 
$\simeq 0.25~{\rm cm}^{-1}$, $\Gamma $ ($T=55$~K) $\simeq 0.8~{\rm cm}^{-1}$. 
These date are in
resonable agreement with ordinary values of hard phonon life times and with
observed low-frequency half-width (see figs 3, 4 ).

3.2 Now let us proceed to consideration of Raman spectra in $KTN$ with 15.7\%
$Nb$ which has the transition from cubic to tetragonal phase at $T_{c}=138.6$~K.
The measurements were carried out at several temperatures in vicinity of
$T_{c}$: $T=160$, 150, 146 and 142~K \cite{2}. At all the temperatures the lines
with two maxima were observed, the first being nearby 
$\omega _{1}\simeq 200~{\rm cm}^{-1}$ 
and the second at 
$ \omega _{2}\simeq 220~{\rm cm}^{-1}$. The intensity of
low frequency maximum essentially increased with $T$ lowering, meanwhile the
intensity of high frequency one became very small at $T=142$~K. 
Since nonlinear
parameter $ \alpha _{2}$ increases with temperature approaching to $T_{c}$
\cite{glinchuk97}, nonlinear effects has to be the largest at $T=142$~K. 
Qualitatively
transformation of spectra from two-peak line to one-sharp peak line at nonlinear
parameter increasing is in agreement with theoretical overcasting (see figs.1,
2) It is seen also that increasing of low frequency maximum intensity with $T$
lowering can be the result of $1/\tau $ decreasing.

To be sure that nonlinear effects are really responsible for observed Raman
spectra transformation we checked if the Eqs.(13-15) fitted experimental
spectra. It was shown that Eq.(\ref{15}) fitted the wing 
($\omega \geq 230~{\rm cm}^{-1}$) 
of Raman line at $T=142$~K for $\alpha _2\simeq 0.015~{\rm cm}$, which
lead to $\omega _c\simeq 20~{\rm cm}^{-1}$ (see Eq.~(\ref{14})). 
This value fits pretty
good the distance between two peaks of observed Raman spectra (see fig.~5),
which speaks in favour of nonlinear effects contribution. In fig.~5a we
depicted line shape calculated with the help of Eq.~(\ref{17}) for $\alpha
 _2=0.012~{\rm cm}$ and $\Delta \simeq 17~{\rm cm}^{-1}$. 
 The later quantity was taken from
measured high frequency maximum width. Note, that obtained $\Delta $ value made
it possible to fit dimensionless $\omega /\Delta $ scale with $\omega $ scale in
fig.~5 and later in fig.~6.

The values of homogeneous broadening parameter $1/\tau $ at different T were
calculated in supposition that it defines mainly by reorientational
frequency  of elastic dipole connected with $Nb$. This frequency temperature
dependence  was measured early \cite{antimirova90} and it was described 
by Arrenius law
\begin{equation}
\label{18}
\frac{1}{\tau }=\frac{1}{\tau _{0}}\exp(-U/T)
\end{equation}
\noindent
with U=200 K and $1/\tau _{0} =7\times 10^{9}$~Hz. 
At $T=150$~K Eq.(\ref{13}) leads
to  the value $(\tau \Delta )^{-1}\simeq 0.004$, which was used in
calculation of  the line, depicted in fig.~5a. It is seen from fig.~5 that the
calculation and  observed spectra look like one another.More detailed
comparison of the theory  and experiment was carried out for Raman line
observed at $T=142$~K (see fig.~6).  Theoretical curve was drawn for
aforementioned parameter $\alpha _{2}=0.015$~cm, obtained from line wing
behaviour and $1/\tau $ was calculated with the  help of Eq.(\ref{18}). It
is seen that theory fits pretty good observed Raman  spectra. This gave
evidence that measured Raman line asymmetry and rapid  change in line shape
when approaching the transition from above is really  connected with
nonlinear effects in $KTN$ with 15.7\% $Nb$. Theory overcasts  strong
increasing of low frequency peak at the region 
$142~{\rm K} > T > T_c$ because  of 
$\alpha _{2}$ increasing and $1/\tau $ decreasing in supposition that there
is no another even small temperature independent contribution to $1/\tau $.

Comparing the obtained data for $KTL$ and $KTN$ one can see that because the
reorientation rate of $Nb$ dipoles is much larger than that of $Li$, the
homogeneous broadening of Raman line in $KTL$ is defined by hard phonon dynamic
whereas in $KTN$ - by $Nb$ elastic moment reorintation. Note that in \cite{2,3}
the reorientation of $Nb$ electric dipole moment were supposed to be the origin
of the homogeneous broadening. However the parameters of the electric dipoles
orientations obtained from fitting of the theory with the experiment were
strongly different from those obtained early in independent measurements
\cite{hochli}.

\section{Discussion}

The proposed theoretical description of Raman spectra shapes  was
perfomed without calculation of correlation function of quasy-static fluctuation
of polarization. The comparison with the caculations of Raman spectra based on
calculation of this correlation function shows that the parameter $\alpha
_{2}\Delta $ corresponds to $v^{2}_{h}/\omega ^{2}_{0}r^{2}_{c}$ where $v_{h}$,
$\Omega _{0}$ and $r_{c}$  are, respectively, a sound velosity, hard mode
frequency at $q=0$ and the correlation length of the polarisation in the pure
lattice. These physical quantities define temperature and concentrational
dependence of obtained values of $\alpha _{2}\Delta $. The value of parameter
$R_{c}/r_{c}$ ($R_{c}$ is correlation radius of the lattice with impurities)
defines the ratio of inhomogeneous and homogeneous contributions and at
$T\rightarrow T_{c}$ ($R_{c}\rightarrow \infty $ ) the line becomes completely
inhomogeneous \cite{3} .  In our approach the parameter of nonlinearity
strongly increases with $T\rightarrow T_{c}$ and becomes much greater than
homogeneous contribution which tends to decrease the magnitude of line maximum.
Therefore qualititavely our results are in agreement with those obtained in
\cite{3}  .

However comparing the frequency dependence of $J(\omega )$ at large $\omega $
 one can see that it was described as $(\omega -\Omega _0)^{-3/2}$ \cite{3}
 whereas in our theory as $(1+4\alpha _{2}\omega )^{-1/2}$. More accurate
 measurements of line intensity decay could be desirable for both theory
 comparison. Moreover, our theory gives the line shape with two maxima for
 intermidiate values of $\alpha _{2}\Delta $, which looks like that observed in
 $KTN$ (see fig.~5). Thus the observed line shape with two maxima naturally
 appeares in our theory whereas the second maximum was not obtained in the
  previous theoretical description \cite{2}.
 Its origin was supposed to be some forbidden transition related to the mixture
 of acoustic and optic modes.

Therefore peculiarities of Raman spectra shape were explained by
proposed theory. In particular it was shown that nonlinear effects leads to
strong assymetry of the lines and to appearing of the low frequency  sharp
maximum, its left hand side is defined by dynamic properties of  the
system. The division between dynamic and static characteristics  contribution
into Raman spectra shape makes  possible to investigate  separately the both
aforementioned characteristics by Raman spectroscopy method.

\section*{Figure captions}
\noindent
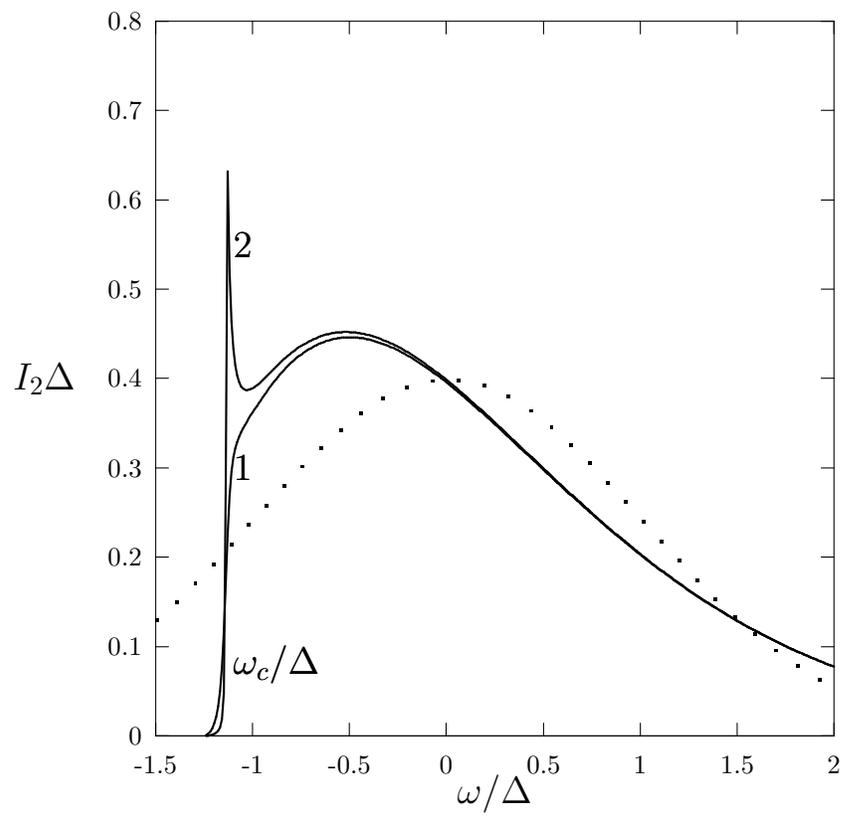
{\bf Figure 1.}
Line shape calculated on the base of Eq.(9) for $\alpha _{2}\Delta $
 =0.22 and $(\tau \Delta )^{-1}$= 0.3 (curve 1), 0.01 (curve 2);
doted line is Gaussian form.

\medskip

\noindent
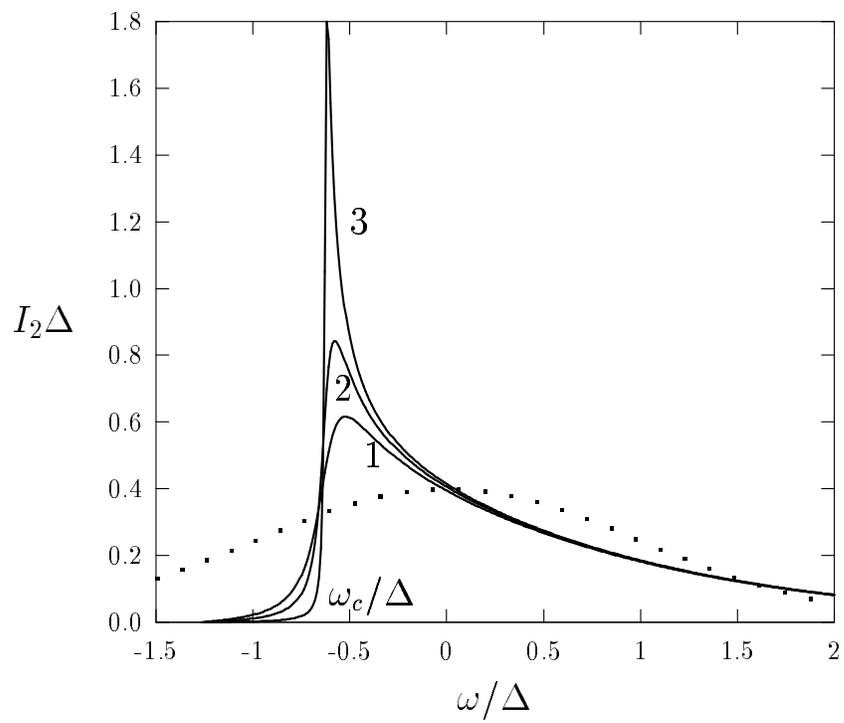
{\bf Figure 2.}
Line shape calculated on the base of Eq.(9) for $\alpha _{2}\Delta $
 =0.4 and $(\tau \Delta )^{-1}$ = 0.1 (curve 1), 0.05 (curve 2), 0.01 (curve
3); doted line is Gaussian form.

\medskip

\noindent
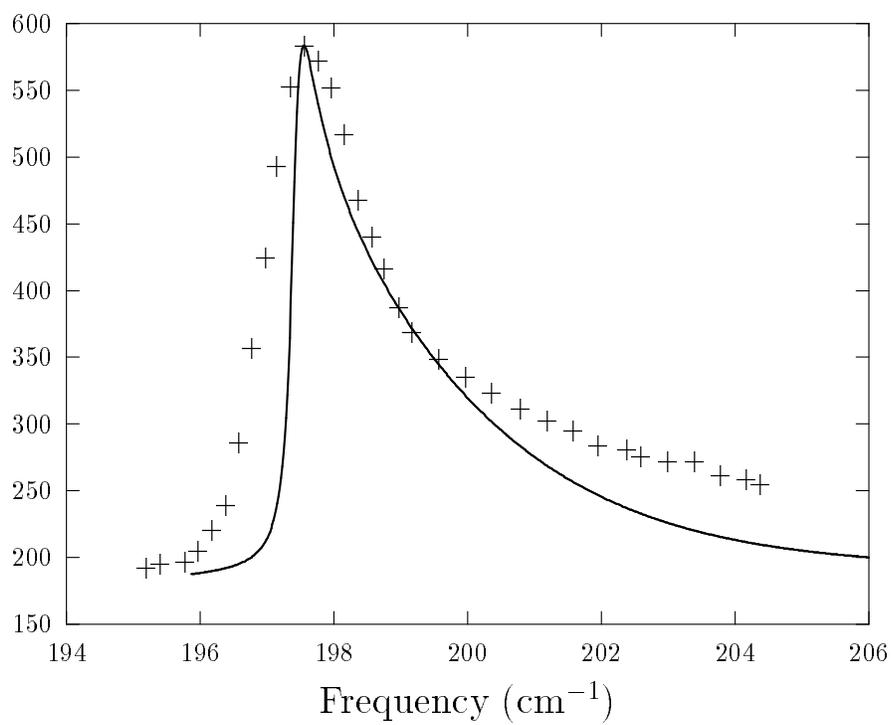
{\bf Figure 3.}
FOR scattering line shape, solid line - theory
at $\alpha _{2}\Delta $ =0.38 and
 $(\tau \Delta )^{-1}$= 0.06, crosses - experimental data
for KLT with 1\% of Li at $T=10$~K [3].
Intensity is represented in arbitrary units.

\medskip

\noindent
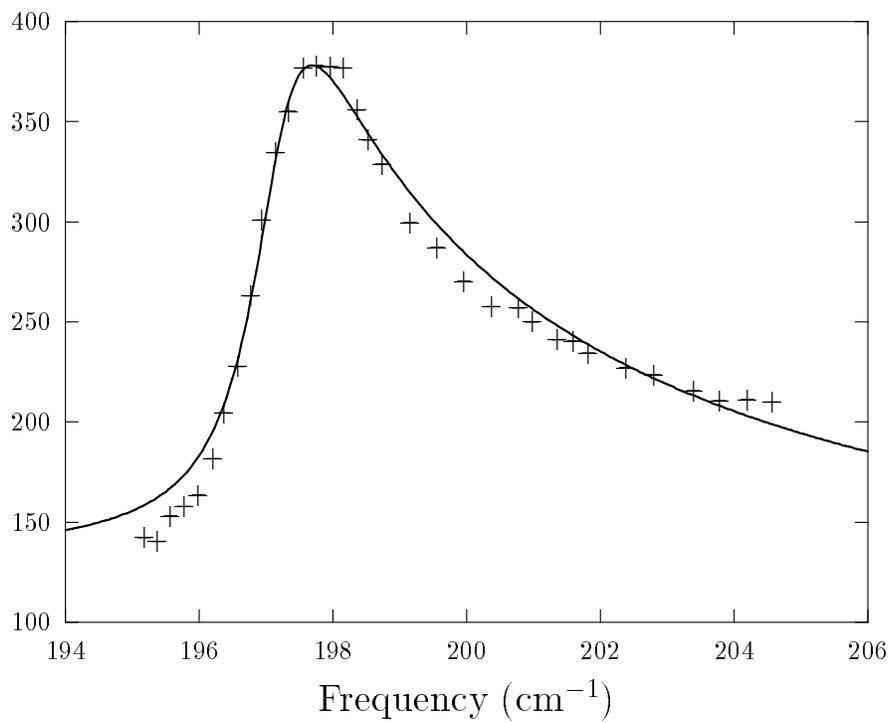
{\bf Figure 4.}
FOR scattering line shape, solid line - theory
at $\alpha _{2}\Delta $ =0.55 and
 $(\tau \Delta )^{-1}$= 0.15, crosses - experimental data
for KLT with 4\% of Li at $T=55$~K [3].
Intensity is represented in arbitrary units.

\medskip

\noindent
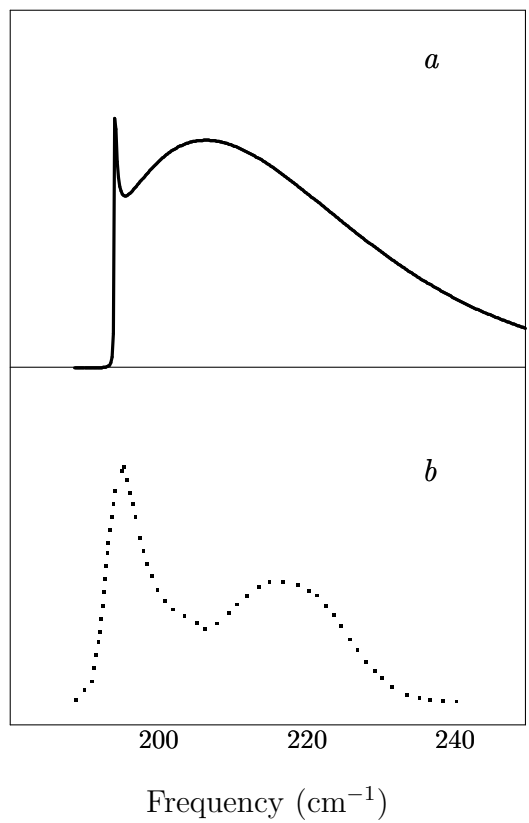
{\bf Figure 5.}
FOR scattering line shape of KTN with 15.7\% of Nb at $T=150$~K [2]
 ({\it b}); calculated line shape for
 $\alpha _{2}\Delta $ =0.21, $(\tau \Delta )^{-1}$= 0.004 ({\it a}).

\medskip

\noindent
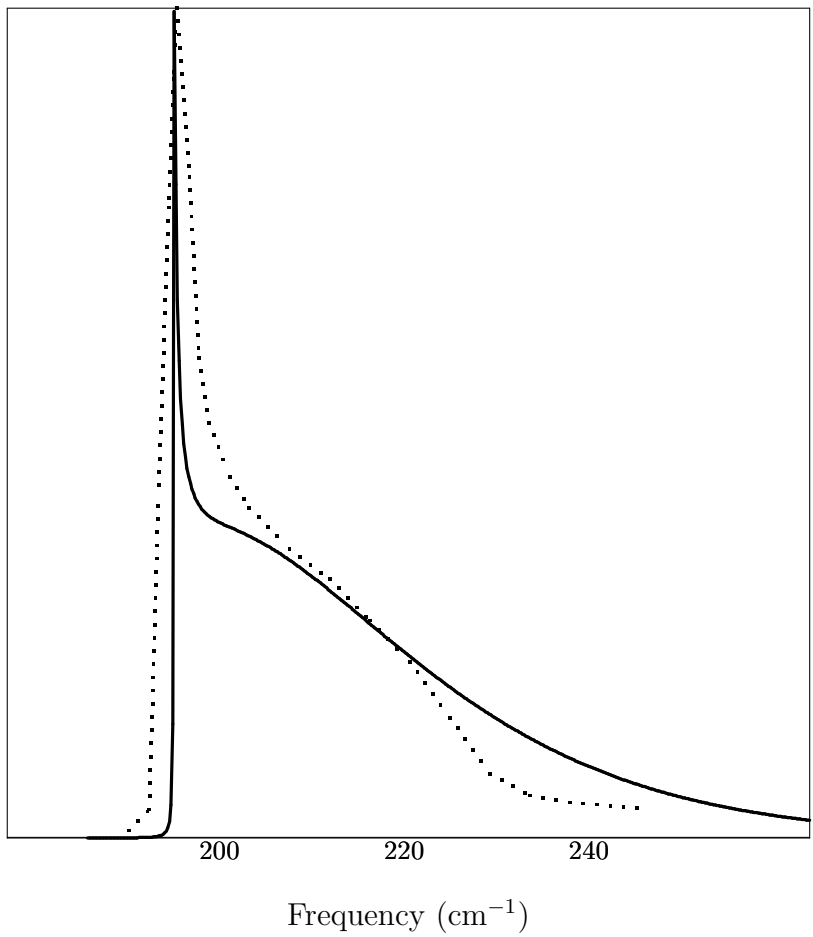
{\bf Figure 6.}
FOR scattering line shape, solid line - theory at
$\alpha _{2}\Delta $ =0.25 and
$(\tau \Delta )^{-1}$= 0.0036, doted line -
for KTN with 15.7\% of Nb
at $T=142$~K [2].

\end{document}